\documentclass[12pt]{article}
\usepackage{hyperref}
\usepackage[usenames, dvipsnames]{color}
\usepackage{enumitem}
\usepackage{amssymb,graphicx}
\usepackage{cite}

\newcommand{\be}{\begin{equation}}
\newcommand{\ee}{\end{equation}}
\newcommand{\bea}{\begin{eqnarray}}
\newcommand{\eea}{\end{eqnarray}}

\begin{document}
\begin{titlepage}

\begin{flushright}
{\today}
\end{flushright}
\vspace{1in}

\begin{center}
\Large
{\bf Froissart Bound, Diffraction Scattering of Hadrons and Scaling at Asymptotic Energies   }
\end{center}

\vspace{.2in}

\normalsize

\begin{center}
{ Aruna Kumar Nayak and Jnanadeva Maharana 
\footnote{nayak$@$iopb.res.in;maharana$@$iopb.res.in }} 
\end{center}
 
\normalsize

\begin{center}
 {\em Institute of Physics \\
Bhubaneswar - 751005  \\    }

\end{center}

\vspace{.2in}

\baselineskip=24pt

\begin{abstract}
The Froissart bound on the total cross section, $\sigma_t$,
 is subjected to test against  
 very high energy data.  We have found no clear 
evidence for its violation. 
 The scaling property of differential  cross section 
in the diffraction region is 
investigated. It exhibits scaling in the ISR, SPS, Tevatron and LHC 
energy domain which had hitherto remained unexplored. 
The slope of the diffraction peak is fitted and the data 
are tested against the rigorous bounds.

\end{abstract}

\vspace{.5in}

\end{titlepage}

\section{Introduction}
The high energy scattering of hadrons is of great interest. The
total cross sections $\sigma^{pp}_t$ and $\sigma^{p\bar p}_t$, from
ISR to LHC energies, 
 are rising and the cosmic ray data continues to follow the  trend. 
The elastic cross sections, $\sigma_{el}$, grow with energy as well.
The high energy 
scattering data are endowed with several fascinating features \cite{pdg_2016}.
 The elastic differential
cross sections exhibit forward peak whose width shrinks as energy grows. 
The slope
of the diffraction peak increases with energy. There are phenomenological
models to study high energy hadronic processes. 
There is another approach to study strong interactions without
appealing to any specific model. It is the S-matrix formulation
poineered by Heisenberg \cite{heisen}. This philosophy,
over the years, has evolved into axiomatic field theory approach as known
at present. The rigorous results derived in this framework are not
based on any model. Indeed, the structure rests on certain fundmantal
axioms to be respected by relativistic quantum field theories (QFT).
The analyticity properties and other attributes of scattering amplitudes
are proved starting from the axioms. The hallmark of the axiomatic
formulation is that the principal results are presented as upper and lower
bounds on experimentally measured quantities. So far there are no
experimental evidences for  the violation of any of these bounds.
One of the most celebrated results is the Froissart-Martin bound on the
total hadronic cross sections, $\sigma_t$. Throughout this paper the high
energy data means the data accumulated in the energy range covered
by accelerators such as ISR, SPS, Tevatron and LHC. Whenever the data
of cosmic rays are incorporated they will be referred to explicitly.
The high energy accelerators have been accumulating data over the years. 
It is worthwhile to focus attention on such data and study afresh how they
put the bounds to tests.

The purpose of this article is to investigate certain areas of strong 
interaction processes which have remained unexplored in the high energy regime so far. 
We shall consider three topics as mentioned below.\\
{\bf 1}. Our intent is to test the validity of the Froissart-Martin bound 
\cite{f1,f2} 
\bea
\label{scale1}
 \sigma_t(s)\le{{4\pi}\over{t_0}}log^{2}\left({s\over{s_0}}\right)
\eea
in the asymptotic energy regime. Here 
 $s$ is the c.m. energy squared, 
$t_0$ is a constant less than or equal to the $t$-channel threshold.
The prefactor in (\ref{scale1}) was fixed by Martin \cite{f2} from
the first principles.  $s_0$ is introduced to make argument of the 
$log^2$ dimensionless and it cannot be determined {\it ab initio}. However,
recently, there has been attempts to determine $s_0$ with additional
inputs \cite{mr}. The first important point to note
is that the total cross section cannot have arbitrary energy dependence. 
Moreover, this bound is not obtained from any model in a perturbative framework,
rather, it is a consequence of general axioms. Therefore, any
experimental evidence for  violation of the
bound (\ref{scale1}) would be a matter of concern. We have adopted a strategy
to test the Froissart-Martin bound in the very high energy regimes 
alluded to earlier. We shall outline our prescription in the next section. \\
{\bf 2.} The scaling of the differential cross section, 
$ {{d\sigma}\over{dt}}(s,t)$, in the diffraction region, i.e. 
in the near forward direction, is an interesting and
intriguing feature in high energy reactions. This attribute has been observed 
in the pre-ISR energy domain in the past. The scaling, in the present context,
is interpreted as follows. The scattering amplitude, 
$F(s,t)$,  is a function of two Lorentz invariant variables $s$ and 
the momentum transferred squared,  $t$.
In the limit, $s\rightarrow\infty$, for fixed physical $t$,  
$F(s,t)\rightarrow f(\tau)$; $f(\tau)$
being a function of a single variable. The mystery about such a scaling
lies in the fact that perturbative QCD techniques cannot be applied to 
explain such an observed scaling. The problem is that in the
kinematical region in question, i.e. large $s$ and small $|t|$, the
reaction is soft; therefore, the power of the  asymptotic freedom 
is inapplicable. We shall elaborate more on this type of scaling in sequel. 
In the past, several models \cite{duane,duane1,duane2,revscale} were proposed
which described high energy scattering and investigated scaling. A good review
of scaling phenomena with extensive references is presented in \cite{revscale}.   We examine scaling 
in the aforementioned energy domain. \\ 
{\bf 3.} The slope of the diffraction peak, $b(s)$, is a very important 
parameter in 
high energy hadronic collisions. It is not only useful to describe
diffraction scattering but also it is utilized by the experimentalists to
extract $\sigma_t$.  
The interests in energy dependence of $b(s)$ is 
 to test the shrinking of the diffraction peak with energy;
furthermore, axiomatic bounds would also be tested.  
Moreover, in the past, the measured
diffraction peak data almost saturated the lower bound for the absorptive
part of differential cross section; leading to the conjecture that the
amplitude is dominated by the imaginary part. 

{\it Remarks:} (i). The analyticity of the 4-point scattering amplitudes
is the cornerstone of the bound (\ref{scale1}). The rigorous results such 
as analyticity and crossing symmetry  
 are proved from axiomatic field theories~\cite{book1,book2,book3,book4}; 
either in the framework
of Lehmann-Symanzyk-Zimmermann formulation~\cite{lsz} or from Wightman's axioms 
\cite{wightman}. We recall the axioms of LSZ for sake of completeness.
{\bf a1.} The states of the system are represented in  a
Hilbert space, ${ \cal H}$. All the physical observables are self-adjoint
operators in the Hilbert space, ${\cal H}$.
{\bf a2.} The theory is invariant under inhomogeneous Lorentz transformations.
{\bf a3.} The energy-momentum of the states are defined. It follows from the
requirements of  Lorentz  and translation invariance that
we can construct a representation of the
orthochronous  Lorentz group. The representation
corresponds to unitary operators, $ U( a, \Lambda)$,  and the theory is
invariant
under these transformations. Thus there are hermitian operators corresponding
to spacetime translations, denoted as $ P_{\mu}$, with $ \mu=0,1,2,3$ 
which have following
properties:
\be
\bigg[ P_{\mu},  P_{\nu} \bigg]=0
\ee
If ${\cal F}(x)$ is any Heisenberg operator then its commutator with 
$ P_{\mu}$
is
\be
\bigg[ P_{\mu}, {\cal F}(x) \bigg]=i\partial_{\mu}{\cal F}( x)
\ee
It is assumed that the operator does not explicitly depend on spacetime
coordinates.
If we choose a representation where the translation operators, $ P_{\mu}$,
are diagonal and the basis vectors $| p,\alpha>$  span the Hilbert
space,
${\cal H}$,
\be
 P_{\mu}| p,\alpha>~=~ p_{\mu}| p,\alpha>
\ee
then we are in a position to make more precise statements: \\
           ${\bullet}$ Existence of the vacuum: there is a unique invariant 
vacuum state
$|0>$ which has the property
\be
 U( a,\Lambda)|0>~=~|0>
\ee
The vacuum is unique and is Poincar\'e invariant.\\
${\bullet}$ The eigenvalue of $ P_{\mu}$, $ p_{\mu}$,
is light-like, with $ p_0>0$.
We are discussing  only the case of   massive states. If we implement
infinitesimal Poincar\'e transformation on the vacuum state then
\be
 P_{\mu}|0>~=~0,~~~ {\rm and}~~~ M_{\mu\nu}|0>~=~0
\ee
from above postulates; note that $ M_{\mu\nu}$ are the generators of Lorentz
transformations.\\
{\bf a4.} The locality of theory implies that a (bosonic) local operator
at spacetime point
$ x^{\mu}$ commutes with another (bosonic)
local operator at $ x'^{\mu}$ when  their
separation is spacelike i.e. if $( x- x')^2<0$. 
Our Minkowski metric convention is $(1,-1,-1,-1)$.
The fixed-$t$ dispersion relations for, $F(s,t)$,  is proved
from the above axiomatic theories and  $t$ must lie within 
the Lehmann-Martin
ellipse \cite{lehmann,martin}.
Moreover, there are several theorems, presented  as upper and
lower bounds  which have stood  the experimental tests so far
 \cite{eden1,eden2,sashanka}.\\
 (ii). The aforementioned
 scaling lies in a kinematical domain (large $s$ and small $|t|$)
such that the perturbative QCD techniques cannot be applied to 
 explain it. 
In the context of this scaling,  the axiomatic 
field theory provides
a precise and rigorous definition and proves the {\it raison de etre} 
 of the scaling \cite{akm}. We have chosen two different
types of scaling variables to investigate how the data respond to our
propositions. The choice is motivated under certain mathematical grounds
and it is not of purely phenomenological as we shall
explain later.\\
  (iii).
The combined data of $\sigma^{pp}_t$ and $\sigma^{p\bar p}_t$ are used 
to test the validity of Froissart bound. 
The point to note is that the rising of $\sigma^{pp}_t$ and 
$\sigma^{p\bar p}_t$ was first observed at ISR. 
It was noted that  
$\Delta\sigma=\sigma^{pp}_t-\sigma^{p\bar p}_t$ is small and
 it decreases with growing ISR energy. The Pomeranchuk's theorem
\cite{pom,martinpom,kinoshita} provides a clue for us from the above observed feature.  
The original theorem stated
that the particle-particle and the particle-antiparticle total  cross
sections would tend to  equal values asymptotically; however, an important
assumption was   
 that \cite{pom} 
 total cross sections
attain constant values at asymptotic energies. We recall that  
$\sigma^{pp}_t$ and $\sigma^{p\bar p}_t$ continue to rise at ISR, SPS, 
Tevatron and LHC energies; cosmic ray confirms the same trend. 
 Note  that the Pomeranchuk theorem is
not a consequence of axioms of the QFT. The generalized theorem
for the case of rising cross sections has been proved \cite{martinrev}.  
We recall that only
 ISR has measured the two cross sections, $\sigma^{pp}_t$ and 
$\sigma^{p\bar p}_t$, at the same energy and therefore, the theorem
can be tested there. When fits to $pp$ and $p\bar p$ cross section
data are extrapolated,  the two curves show a tendency to converge
asymptotically. We justify, therefore, to combine the two sets of data
and fit the cross sections.  

The article is organized as follows. The next section, Section~\ref{sec2}, is
devoted to study of phenomenae mentioned earlier. 
In the section~\ref{sec2a}, we propose a formula
to fit the combined total cross section data, $\sigma^{pp}$ and 
$\sigma^{p\bar p}_t$, from ISR energies all the way up to the cosmic rays.
We present a justification for the choice of the fitting formula. The
study of the scaling of the elastic differential cross section is 
carried out in section~\ref{sec2b}. We briefly recall the essential results on
this type of scaling envisaged from rigorous mathematical structure.
The results from the axiomatic approach are utilized for our purpose.
We examine the scaling of diffractive elastic scattering.
We choose two different scaling variables to investigate this phenomena. 
Section~\ref{sec2c} is devoted to the study of the slope of
the diffraction peak, $b(s)$. As we have noted earlier, it is an important
experimentally measured parameter. We mention in passing that, to our
surprize, there has not been much activity to explore the energy
dependence of $b(s)$. We fit $b(s)$ from ISR energies to the LHC energy.
Moreover, there are rigorous bounds on the slope of the absoptive diffraction
diffrential cross sections, $b^A(s)$. There is no strong reason that why
the data should respect these bounds. The motivation lies in the fact that,
  in the past, the data seemed to saturate the lower bound which led to
an interesting conjecture. We test the bounds against the data. The third
section contains the summary of our results and conclusions.

\section{Analysis of Experimental Data}
\label{sec2}
We proceed to fit $\sigma_t$, investigate the scaling of differential cross sections, and 
theenergy dependence of the slope parameter. 

\subsection{The fit to Total Cross Sections}
\label{sec2a}
The purpose is to fit the total cross section data in order to test the 
Froissart-Martin bound. We would like to draw attention to the following facts.
The PDG \cite{pdg_2016} has presented a very good fit to total cross sections
of a large number of hadronic reactions over a wide range of energies. Let us
discuss their fit to $\sigma^{pp}_t$. They considered data from 5 GeV all
the way upto LHC energies; moreover, the cosmic ray data are also 
incorporated in the fit. We recall that the Froissart-bound saturating
energy dependence is incorporated in their fit. In other words, 
the term is $log^2{s\over{s_0}}$. The coefficient of this term, the 
Heisenberg constant, $H$, and $s_0$ are treated as floating parameters. In 
addition there is a constant term, P, associated with Pomeranchuk trajectory and
several other terms corresponding to the contributions of Regge trajectories were added.
It is important to note that the total cross sections are flat from 5 GeV
till the pre-ISR energy domain and there are vast data points measured
with very good precision. Moreover, in the energy interval noted above, 
the Regge trajectory, such as $\rho$, $A_2$.. contribitions are important
although they fall off as power of $s$ compared to the Pomeranchuk term and
the Heisenberg term. In the fit to $\sigma_t$, the vast set of data points
where $\sigma_t$ is constant, terms subdominant compared to $log^2s$ term,
play a crucial role. Moreover, the fit does not set out to test the
Froissart-Martin bound since its exponent is fixed as $2$. Our objective is to
test the bound; as a consequence, we take the power of the exponent to be
a floating parameter.   
A point is in order. 
It was claimed, in the past,  in a fit to $\sigma^{pp}_t$, that there is
evidence for violation of the Froissart bound \cite{menon}. 
Subsequently this claim was refuted by Block and Halzen \cite{halzen}.
Note  that  those two papers covered 
  the same energy range as was covered  by the PDG \cite{pdg_2016} i.e.
from 5 ${\mathrm GeV}$ to cosmic ray energies.
There is a recent comprehensive review \cite{yogi} on
high energy collisions which covers diverse aspects of  hadronic
processes and we refer it to the interested reader. 

Let us discuss our proposal to fit the total cross section data.
We choose the following parameterization to fit the combined data.
\bea
\label{ourfit}
\sigma_t=H log^{\alpha}\left({s\over{s_o}}\right)+P
\eea

$H$ and $P$ are the Heisenberg and Pomeranchuk constants, respectively. 
P is the contribution of the Pomeranchuk
trajectory in the Regge pole parlance. 
The constants $H$, $P$ and 
 $\alpha$, are  free parameters and  are 
determined from the fits.
 We fix $s_0=16.00 ~{\mathrm GeV}^2$, taking a hint from the PDG fit.  
PDG adopted the following strategy to fit $\sigma_t$ data.
For the fit to $\sigma^{pp}_t$ the chosen energy range was  
from 5 ${\mathrm GeV}$ to cosmic ray regime. 
The Froissart-bound-saturating energy dependence is incorporated in their 
fitting procedure.
Note that in the pre-ISR energy regime the measured cross sections are flat 
and measured with very good precision. 
Moreover,  Regge pole
contributions, with subleading power behaviors,
 should be included in the pre-ISR energy domain.
However, in the energy
range starting from ISR, the Regge contributions are negligible.
It is worth while to discuss and justify the reasons for not including the  
contributions of subleading
Regge poles to $\sigma_t$ in the energy range starting from ISR point and 
beyond (where our interests lie).
Moreover, the subleading Regge poles have an important roles to play
in fitting $\sigma_t$ where the data are collected from the accelerators
of those period. 
 We refer to \cite{eden1} and to the review article
of Leader \cite{leader} for detailed discussions. We recall that $\sigma_t$
is almost constant in the "low energy" (pre-ISR) region, however, 
contribution of the Regge tail is also necessary. Let us consider the case
of $pp$ scattering to get a concrete idea. 
The Pomeranchuk trajectory contributes a 
constant term to $\sigma_t$ and its intercept is $\alpha_P(0)=1$.
Then there are subleading trajectories corresponding to $\omega$, $\rho$,
$A_2$, $\phi$, etc.  When   a fit to $\sigma^{pp}_t$ was considered by 
Rarita et al \cite{roger}, they concluded, from numerical fits, 
that the $\omega$ trajectory dominates \cite{leader,roger} and the contributions
of other Regge trajectories is quite small \cite{rp}. 
They found that the Regge residue (interpreted as the Regge trajectory coupling)
is $R_{pp\omega}\approx 15.5$ mb and $\alpha_{\omega}(0)\approx 0.45$ and 
the Regge scale, to define a dimensionless ratio (say
$s\over{s_*}$) is $s_*=1~GeV^2$. Let us estimate what is the 
contribution of the $\omega$-trajectory to $\sigma^{pp}_t$ at the ISR energy. 
The contribution of the $\omega$-trajectory to $\sigma_t$ is 
quite small  in the
 energy range from ISR to LHC. For example, at ISR energy
of $\sqrt{s}=23.5~{\mathrm GeV}$, 
the $\omega$-Regge pole contribution to $\sigma_t$ is 
approximately  $0.5$ mb  whereas at LHC, 
for $\sqrt{s}=8~{\mathrm TeV}$, it is $\approx 0.001$ mb; 
the corresponding $\sigma_t$ are $\approx 39~mb$ and $\approx 103~mb$ 
at $23.5~{\mathrm GeV}$ and $8~{\mathrm TeV}$ respectively. The 
parametrizations of \cite{roger} is used for the above estimates.
Consequently, for our purpose, the parameterization
(\ref{ourfit}) is well justified. 

\begin{figure}[!htbp]
  \centering
  \includegraphics[width=0.7\textwidth]{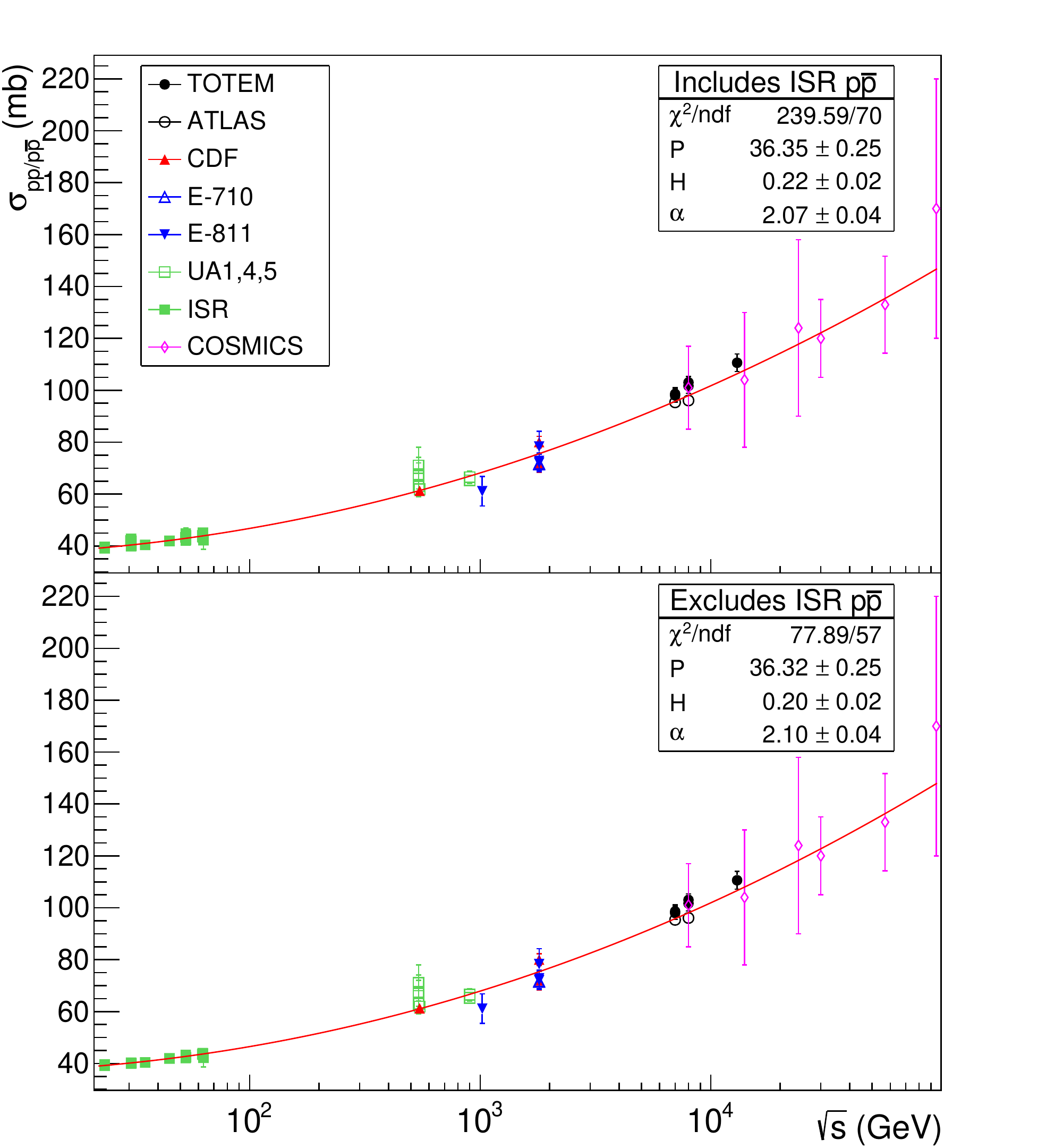}
  \caption{ $\sigma_t^{pp}$ and $\sigma_t^{p\bar{p}}$ against $\sqrt{s}$,
    measured by various experiments.
    The data points are fitted to a function defined in~(\ref{ourfit}). 
    The upper plot includes both $pp$ and $p{\bar p}$ data points,
    while the lower one excludes $p{\bar p}$ data points from ISR experiments.
  }
  \label{fig:sigma_vs_sqrt_s}
\end{figure}

We considered the combined data of  $\sigma^{pp}_t$ and $\sigma^{p\bar p}_t$ for
the energy range as mentioned earlier. 
ISR data are from
$\sqrt{s}~=~23.5$ to 63 ${\mathrm GeV}$
\cite{isr_sigma_t_1, isr_sigma_t_2, isr_sigma_t_3, isr_sigma_t_4,
isr_sigma_t_5, isr_sigma_t_6,
isr_sigma_t_7, isr_sigma_t_8, isr_sigma_t_9, isr_sigma_t_10}.
SPS data are at 540, 541, 546, and 900 ${\mathrm GeV}$ 
\cite{ua1_sigma_t, ua4_sigma_t_540, ua4_sigma_t_540_2,
ua4_sigma_t_541, ua4_sigma_t_546, ua5_sigma_t, ua5_sigma_t_2},
whereas Tevatron data points are at 546 ${\mathrm GeV}$, 1.02 ${\mathrm TeV}$, 
and 1.8 ${\mathrm TeV}$ ~\cite{cdf_sigma_t_546, cdf_sigma_t_1800,
e811_sigma_t_1800, e811_sigma_t_1800_2, e710_sigma_t_1020, e710_sigma_t_1800,
e710_sigma_t_1800_2, e710_sigma_t_1800_3}.
The LHC data points are at 7, 8 and 13 ${\mathrm TeV}$ 
\cite{atlas_sigma_7k, atlas_sigma_t_8k, totem_sigma_7k, 
totem_sigma_7k_2, totem_sigma_7k_3,
totem_sigma_8k, totem_sigma_t_8k_2, totem_sigma_13k}.
Cosmic ray points are at 8, 14, 24, 30, 57, and 95 ${\mathrm TeV}$
\cite{Zbigniew, flyeye_sigma_t, auger_sigma_t, TelescopeArray}.
The measured values of cross sections against $\sqrt{s}$,  
along with the fitted curve, are shown in Fig.~\ref{fig:sigma_vs_sqrt_s}. 
The fitted values for the parameters are $P=36.4\pm0.3~mb$, $H=0.22\pm0.02~mb$, 
and $\alpha=2.07\pm0.04$. 
The quality of the fit, as reflected by the $\chi^2/n.d.f.$ is found 
to be moderate due to inclusion of both $\sigma^{pp}_t$ and $\sigma^{p\bar p}_t$ 
measurements from ISR. A fit excluding $\sigma^{p\bar p}_t$ from ISR, 
as shown in Fig.~\ref{fig:sigma_vs_sqrt_s} (lower), improves the fit quality without 
significantly changing the value of the fit parameters. 
We find no conclusive evidence for the violation of the  Froissart bound.
   
\subsection{Scaling of the Diffraction Cross Section}
\label{sec2b}
The scaling observed in the diffraction region is an unexpected 
feature.  This scaling phenomena is mysterius in the sense
that we are unable to explain it from perturbative QCD perspectives.
The kinematical region of interests (i.e. large $s$ and small $|t|$)
does not fall into the purview of the perturbative QCD terrain.
Moreover,
  we draw attention to the fact that so far there is no careful 
analysis of the scaling 
 in the  diffraction region of  very high energy regime which we intend
to explore.
 Therefore, it deserves a thorough investigation. 
A precise definition of scaling and its derivation was provided from 
axiomatic field theory approach.
The necessary ingredients stated and proved from the axiomatic field theory
are \cite{akm}: (i)  The scattering amplitude, $F(s,t)$, is holomorphic in
the disk $|t|<t_0$ for
any $s$ in cut s plane, where $t_0$ is a constant less than or equal to the
$t$-channel
threshold. (ii) The polynomial boundedness:  $|F(s,t)|$ is bounded by $s^N$,
$N\in Z$ and N is finite,  for $|t|<t_0$ and
$s\rightarrow +\infty$. (iii) $F(s,t)$ satisfies unitarity in the s-channel.
It follows from the results (i)-(iii) that $F(s,t)$ satisfies the bound
\bea
|F(s,t)|\le \left[{{4\pi\sigma_{el}(s)}\over{t_0}}\right]^{1/2}slogs~
exp \left[\left({|t|}\over{t_0}\right)^{1/2}logs \right]
\eea

The Pomeranchuk-theorem-violating amplitude (see \cite{akm})
asymptotically behaves as
\bea
F(s,0)\sim i\sigma s+Cs~logs, ~~0<|C|<(4\pi\sigma/{t_0})^{1/2}
\eea
where $\sigma$ is the total cross section. There is  a more refined
mathematical statement
on the behavior of $F(s,0)$ in \cite{akm}. 
A general class of amplitudes  of the form 
\bea
\label{frsatu}
{F(s,0)\sim(i\alpha+\beta)s(log s)^2}
\eea
was considered by \cite{akm}; where
 $0<\alpha\le({{4\pi}\over{t_0}}),~|\beta|\le{({{4\pi\alpha}\over{t_0}})^{1/2}}$ and $F(s,0)$ respects the Froissart bound.
Our sole interest is to recapitulate the scaling property succinctly derived 
and proved in \cite{akm}.  A function
\bea
f(s,\tau)={{F(s,-t_0\tau(log s)^{-2})}\over F(s,0)}
\eea
was introduced in \cite{akm}.
The scaling  is interpreted  in the following sense:
define  $f(\tau)=lim_{s\rightarrow\infty}f(s,\tau)$. It was proved
that $f(\tau)$ is an entire function with following properties
\bea
f(0)=1,~~
|f(\tau)|< \left({{4\pi C_0}\over{t_0}}\right)^{(1/2)}e^{\sqrt\tau},~{\rm  for~all~ \tau}
\eea
where $C_0={{\alpha}\over{\alpha^2+\beta^2}}$;  $\alpha$ and
$\beta$ are as in  (\ref{frsatu}). In nutshell,  it summarizes the
scaling phenomena as derived  from the axiomatic perspectives.

We have chosen two different types of scaling variables based on certain
 theoretical consideration.
(a) The first choice is $\tau_1={{1\over{16\pi}}}{{\sigma_t}^2\over{\sigma_{el}}}t$. 
The technique of group contraction 
was adopted to describe high energy diffraction scattering \cite{spmjm}.
It was demonstrated that the  scaling occurs in the small $|t|$ domain
at high energies. 
The high energy collision of spinless particles, in small angles, was envisaged. 
 Let us  consider a sphere of radius $\gamma$ with polar angles 
$\theta$ and $\phi$. The amplitude, for scattering of scalars,  depends only on 
$\theta$. Let us  
  focus on scattering in the forward regions such 
that $\gamma\theta\rightarrow finite$ 
as $\theta\rightarrow 0$ and $\gamma\rightarrow\infty$. The group contraction
technique 
  \cite{jtalman} was invoked as follows.
  $O(3)$ is the little group of the Lorentz group $O(3,1)$. In the limit, 
$\gamma\rightarrow\infty$,   $O(3)$
   contracts to  $E_2$, the group of translations.  
Let $L_i, i=1,2,3$
  be  generators of $O(3)$ satisfying $[L_i,L_j]=i\epsilon_{ijk}L_k$.  
Define $L_2=\gamma Q_2$ and $L_1=-\gamma Q_1$ with $L_3$ unaltered. The
  angular momentum algebra assumes the form 
$[L_3,Q_2]=iQ_1$, $[L_3,Q_1]=-iQ_2$  and  $[Q_1,Q_2]=0$ in the limit 
$\gamma\rightarrow\infty$.
  Notice that  $Q_1, Q_2$ are translation operators of 
$E_2$. $L_3$ generates rotation around the 3rd axis. If $q_a$ and $q_b$ are 
the momenta defining  
a representation
  of $E_2$, then the eigenvalue of Casimir of 
$O(3)$, $l(l+1) \rightarrow (l+1/2)^2$ in the limit $\gamma\rightarrow\infty$.
The partial wave amplitude, $f_l(s)\rightarrow f(q,s)$
  where $q=\sqrt{(q_1^2+q_2^2)}$. The  expression for the
  scattering amplitude \cite{spmjm} is derived from
the following ingredients: (i)  Partial wave unitarity. 
(ii) Optical theorem to relate
the absorptive forward scattering amplitude to $\sigma_t$. 
And (iii) utilize the standard relation
  $\sigma_{el}={{4\pi}\over{p^2}}\sum_0^{\infty}(2l+1)|f_l(s)|^2$, 
$p$ being the c.m. momentum. Next step is to replace $f_l(s)$ 
by $f(q,s)$ and the sum over $l$ goes over to an integral in $q$. 
Moreover, $f(q,s)$ is assumed to have
  a Gaussian distribution in $q$. Finally, the normalized differential 
cross section, after a few algebraic steps,
was derived to be 
  \bea
  \label{scaling1}
  {{d\sigma}\over{dt}}={{d\sigma}\over{dt}}|_{t=0}
e^{t{{\sigma_t^2}\over{16\pi\sigma_{el}}}}  
\eea
The identification of $\tau_1=|t|{{\sigma_t^2}\over{16\pi\sigma_{el}}}$ 
as a scaling variable was noted in \cite{spmjm}.
Indeed the diffraction scattering data, available at that juncture,  
for $\pi^{\pm}p$, $k^{\pm}p$, $pp$ and $p\bar{p}$ over a wide energy 
range exhibited the scaling behavior. The objective is to examine
whether the present data exhibits the scaling.\\
(b) The other choice  is $\tau_2=|t|b(s)$,
where $b(s)$ is the experimentally measured slope parameter. 
Cornille and Martin~\cite{cm} have demonstrated rigorously the existence 
of this scaling 
variable.
The data prior to SPS, Tevatron, LHC era
exhibited the scaling \cite{cm}.
Subsequent works \cite{mahoux,ar,jm,jmjkm}  
investigated and tested the scaling with the vast data of that era.
Notice that the data in the forward direction is conveniently
parameterized  as
\bea
 \label{bt}
 {{d\sigma}\over{dt}}(s,t)={{d\sigma}\over{dt}}|_{t=0}e^{B(s,t)} 
\eea
 where $B(s,t)$ is usually  expanded in a power series in $t$, around $t=0$, as
$B(s,t)=b(s)t+c(s)t^2+...$; where $b(s)$ is the slope of the diffraction peak,
$c(s)$ is the
 curvature and so on. Thus the $b(s)t$ term dominates in the small
t, near forward region, while fitting ${{d\sigma}\over{dt}}$.
We restrict the values of $|t|$ to the range that is used in
the corresponding measurements to extract the values of $b(s)$ 
assuming a purely exponential dependence of ${d\sigma}\over{dt}$ on $t$.
If we include larger values of $|t|$ then the data favor the
retaintion of the higher curvature
 terms in the expansion of $B(s,t)$ in  (\ref{bt}). 

\begin{figure}[!htbp]
  \centering
  \includegraphics[width=0.65\textwidth]{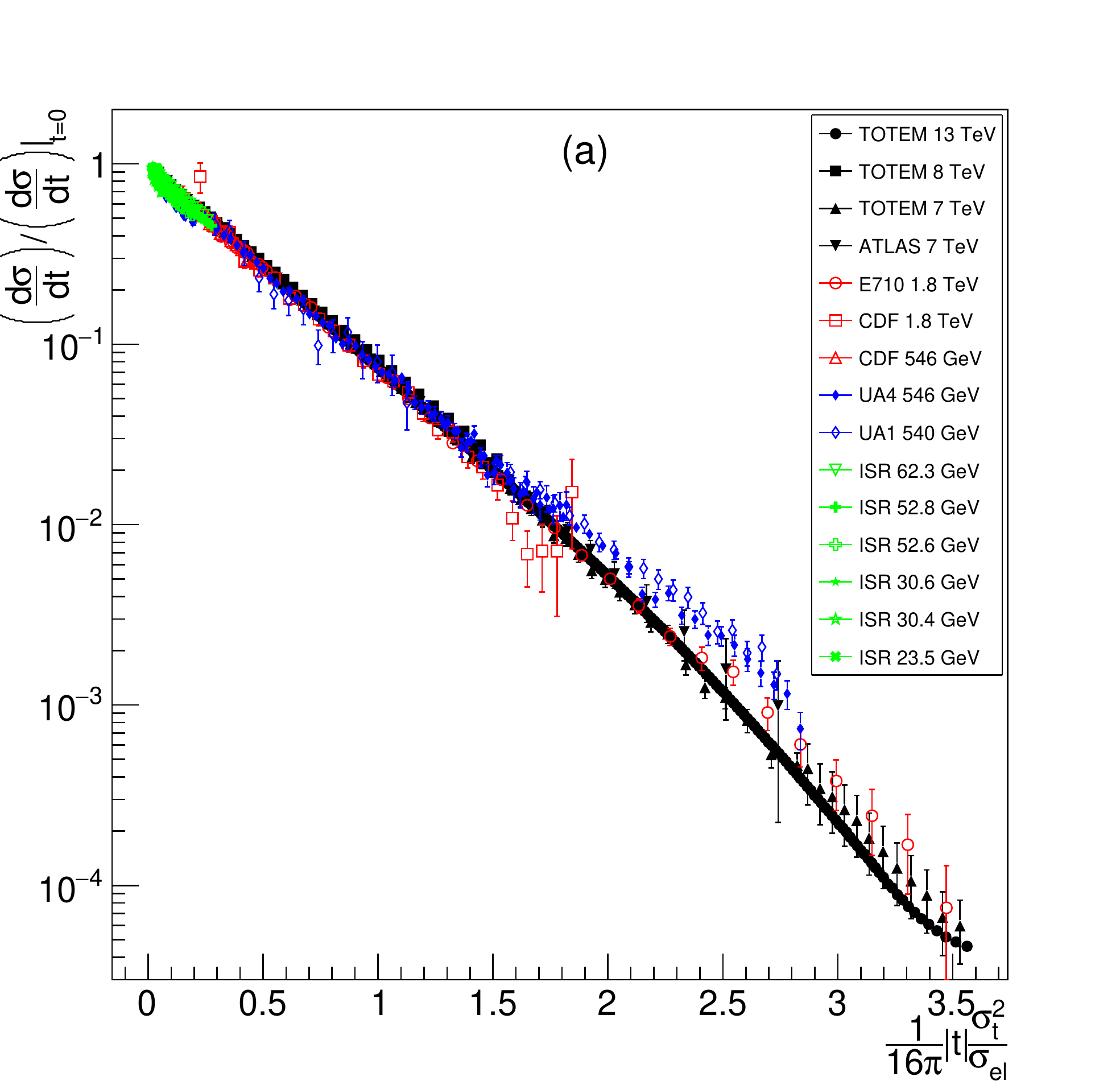} \hfill
  \includegraphics[width=0.65\textwidth]{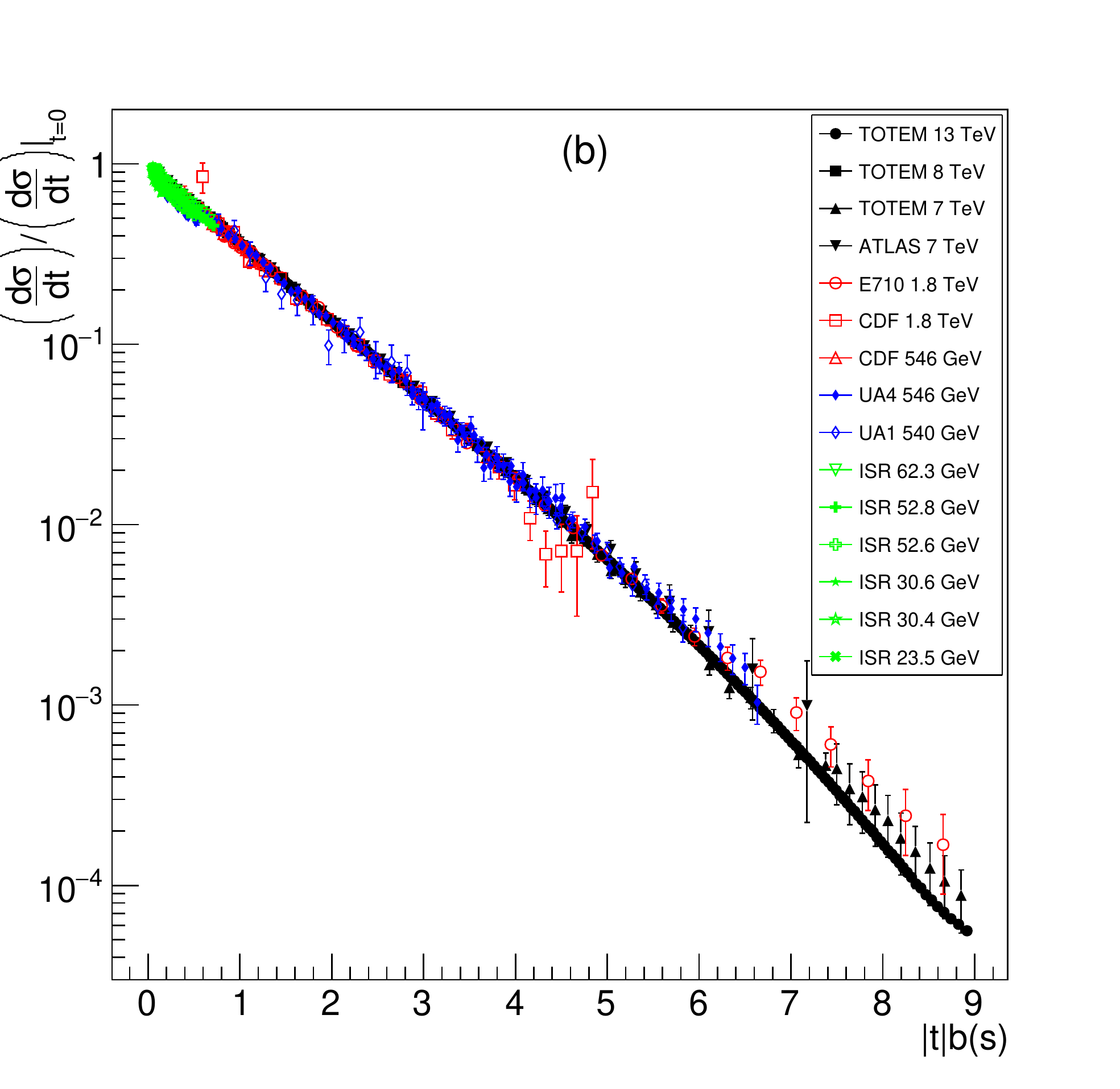}
  \caption{Distributions of the ratio ${{d\sigma}\over{dt}}(s,t)/{{d\sigma}\over{dt}}|_{t=0}$ as function of (a) ${{1\over{16\pi}}}|t|{{\sigma_t}^2\over{\sigma_{el}}}$ and (b) $|t|b(s)$, measured by various experiments for $pp$ and $p{\bar p}$ collisions.
  }
  \label{fig:dsigmaOverdt}
\end{figure}

We have displayed scaling in the
diffraction region for the choice
$\tau_1={{|t|\sigma_t^2}\over{16\pi\sigma_{el}}}$ and $\tau_2= |t|b(s)$, 
 in Fig.~\ref{fig:dsigmaOverdt}. We use $|t|$ in our 
definition
of scaling variables for convenience ( t is always negative in the physical 
region for s-channel reactions).
We have not included a set of ISR data points in a very small $|t|$
region; this point is discussed in detail in the context of extraction of 
$b(s)$ in sequel. 
If those points are included in the  scaling plot
they deviate from  the scaling displayed by rest of the data. 
It is important to note that the
rest of the data exhibits scaling so nicely from ISR to LHC energy range. 
Now let us compare and contrast the scaling phenomena with the two
scaling variables.
For the purpose of examining the scaling for 
the measured values of ${d\sigma}/{dt}$
as function of $|t|$ data are taken from
\cite{ua1_sigma_t, atlas_sigma_7k, totem_sigma_7k, isr_Amos1985, ua4_dsigma_1, 
ua4_dsigma_2, ua4_dsigma_4, cdf_sigma_el, e710_dsigma, totem_dsigma_8k, 
totem_dsigma_13k}.
We analyze the scaling property with
$\tau_1={{1\over{16\pi}}}{{\sigma_t}^2\over{\sigma_{el}}}|t|$ as the scaling
variable and 
the covered range of $\tau_1$ is:  $0.02\le \tau_1\le 3.6$. The scaling is very
good for the case of $\tau_1$  for small values of $|t|$ as  depicted in 
Fig.~\ref{fig:dsigmaOverdt}(a). Note that the scaling is excellent for  the 
 $\tau_2$ variable in the small $|t|$ region and 
 $0.05\le \tau_2\le 9.0$, which is displayed in Fig.~\ref{fig:dsigmaOverdt}(b).
There is a study of scaling in hadronic scattering
\cite{russian}; however, their approach is different and their
main focus is not in the kinematical domain of our  interests.

\subsection{The Slope of the Diffraction Peak}
\label{sec2c}
\begin{figure}[!htbp]
  \centering
  \includegraphics[width=0.7\textwidth]{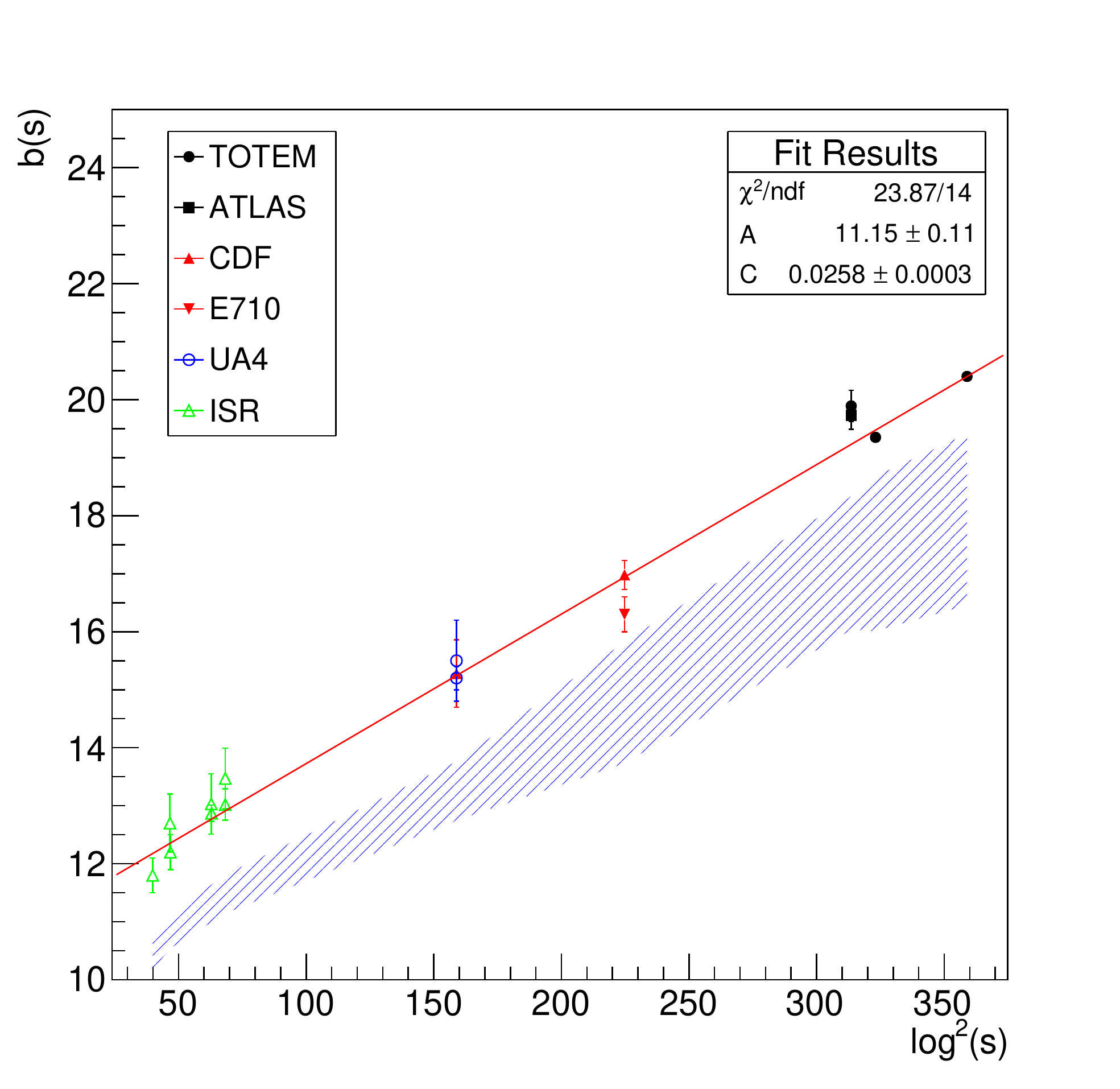}
  \caption{Plot of measured value of $b(s)$ as function of $log^{2}(s)$ as obtained from various 
    experiments. The error band is the lower limit as defined in~(\ref{bound_on_b}).
  }
  \label{fig:b_vs_logsSquare}
\end{figure}

The slope of the diffraction peak is defined to be
  \bea
b(s)={d\over{dt}} log \left({{d\sigma}\over{dt}}\right)|_{t=0}
\eea
and it is experimentally extracted from diffraction scattering data.
Another object of interests is the slope of absorptive differential cross 
section
\bea
b^{A}(s)=  {d\over{dt}} log \left({{d\sigma^A}\over{dt}}\right)|_{t=0}
\eea
There are rigorous  lower and upper bounds on   $b^A(s)$ .
\bea
\label{bound_on_b}
{2\over 9}\left[{\sigma_t^2\over{4\pi\sigma_{el}}}-{1\over{k^2}}\right]\le b^A(s)\le
{1\over{2(t_0-\epsilon)}}\left[log \left({s\over{\sigma_t}}\right)\right]^2
\eea
 The lower bound is due to  Martin and MacDowell  \cite{mm} whereas 
the upper bound was proved by Singh \cite{singh}.
 We have fitted the slope with the
parameterization $b(s)= A+C log^2({{s\over s'}})$ and chose
 $s'=1 ~{\mathrm GeV}^2$ so that the argument of the $log$ is dimensionless.
The fit is depicted  
in Fig.~\ref{fig:b_vs_logsSquare}. Note the following:
the measurement of ${{d\sigma}\over{dt}}(s,t)$ exhibits a sharp
rise in the range $0.0004~\le~|t|~\le~0.004$. In this
 $|t|$ interval, ${{d\sigma}\over{dt}}$
 cannot be fitted by parameterizing it with only a $b(s)$.
 The experiment \cite{isr_Amos1985}  fits the data by adding the proton 
form factor
 term to account for the coulomb effect. Block and collaborators \cite{block}
consider an eikonal model to study pp scattering. They argue that in order
to explain behavior of ${{d\sigma}\over{dt}}$ in the said $|t|$ region, the
effects of higher curvature terms be considered. We leave aside this set of
data points of the particular ISR experiment to fit $b(s)$
taking into account the values of $b(s)$ as presented in various experiments.
These experiments have extracted $b(s)$ 
 from the fits to ${{d\sigma}\over{dt}}$ in near
forward direction.
Let us examine how the measured $b(s)$ stands up
against the bounds (\ref{bound_on_b}). We remind that
the bounds (\ref{bound_on_b}) are proved for $b^A(s)$. Therefore, the
data need not respect them. When the lower bound \cite{mm} was proved
and  was tested against experiments, the data nearly saturated the bound.
Consequently, it was conjectured that the amplitude is dominantly imaginary~\cite{mm}.
In order to compare (experimentally) measured $b(s)$ against
MacDowell-Martin bound we used measured values of $\sigma_t$ and $\sigma_{el}$.
Thus the bound depicted in Fig.~\ref{fig:b_vs_logsSquare} is a band 
due to the uncertainties
 of $\sigma_t$ and $\sigma_{el}$ in measurements. The data
satisfies the Martin-MacDowell bound, however. We remark that
the upper bound (\ref{bound_on_b}) lies way above the data due to the
presence of the factor $2t_0,~t_0=4m^2_{\pi}$, in the denominator.
The measured values of $b(s)$ are taken from
\cite{isr_Amos1985, ua1_sigma_t, ua4_dsigma_4, cdf_sigma_el, e710_dsigma, 
atlas_sigma_7k, totem_sigma_7k, totem_dsigma_8k, totem_dsigma_13k},
which were used for the fit.

\section{Summary and Conclusions} 
One of our objectives was to test the Froissart bound against experiments.
We selected the set of data points where the total cross sections start 
rising. Since $\Delta\sigma$ is quite small at ISR energy and keeps
diminishing with growing $s$ we justified the inclusion of both $pp$
and $p\bar p$ data in the fitting of $\sigma_t$. 
We set out to verify scaling of
the diffraction data at asymptotic energies  with small $|t|$.
We note that this attribute of elastic scattering at asymptotic
energy has hitherto escaped attentions and scrutiny.
It is quite satisfying that the data  exhibit very good scaling for both
the choices: $\tau_1$ and $\tau_2$.  The slope $b(s)$ is a very important 
parameter in the study of hadronic collisions and  
 is used to extract $\sigma_t$ since the point $t=0$ is not accessed
in  experiments.
We have presented a fit to 
$b(s)$ data. The axiomatic bounds of $b^A(s)$ was compared with the data,
although the data need not necessarily respect them; however, they 
are consistent with the bounds. \\
The following remarks are in order. It is an important issue to ponder
over the violation of the Froissart-Martin bound. So far the experimental
data are consistent with the bound. If, in future, the experiments unambiguously
confirm  violation of the bound then it will be a matter of concern. One
possible explanation would be to question the axioms of local field theories.
However, these field theories do not lead only to the bound on total cross
sections. There are other important inferences drawn from the axiomatic
field theory. For example, the CPT theorem \cite{wightman}
is another important outcome which has been put to tests.
Therefore, it will not be so easy to abandon any of the axioms for local field theories. 
On the other hand, if we accept the axioms of local field theory, then new
physics paradigms might have to be introduced. Thus observation of the
violation of the bound on $\sigma_t$ would lead to very important
consequences. 
In conclusion, we have examined several
interesting and important aspects of very high energy scatterings and
we have utilized the data to explore some of the phenomena which had
not been explored in the recent years. 
We tested the validity of the sacred Froissart bound and it still
stands.


\begin{thebibliography}{100}
\bibitem{pdg_2016} C. Patrignani et al. (Particle Data Group), 
Chin. Phys. C, 40, 100001 (2016).
\bibitem{heisen}  W. Heisenberg, Zeitschrift f\"ur Physik, {\bf 133}, 65 (1952) 
\bibitem{f1} M. Froissart, Phys. Rev. {\bf 123}, 1053 (1961).
\bibitem{f2} A. Martin, Phys. Rev. {\bf 129}, 1432 (1963).
\bibitem{mr} A. Martin and S. M. Roy, Phys. Rev. {\bf D 89}, 045015, 2014.
\bibitem{duane} J. Dias de Deus, Nucl. Phys. {\bf B59}, 231 (1973). 
\bibitem{duane1} A. J. Buras and J. Dias de Deus, Nucl. Phys. {\bf B71}, 481 (1974).
\bibitem{duane2} V. Barger, J. Luthe and R. J. N. Phillips, 
Nucl. Phys. {\bf B88}, 237 (1975).
\bibitem{revscale} P. Valin, Phys. Rep. {\bf 203}, 233 (1991).
\bibitem{book1} S. S. Schweber, Introduction to Relativistic Quantum Field 
Theory, Herper and Row, 1961. 
\bibitem{book2} R. Jost, General Theory of Quantized Fields, 
American Mathematical Society, 1965.
\bibitem{book3} R. F. Streater and A. S. Wightman, CPT Spin Statistics and 
All That, W. A. Benjamin, Inc. New Yorm Amsterdam, 1964. 
\bibitem{book4} C. Itzykson and J.- B. Zuber, Quantum Field Theory,
C. Itzykson and J.-B. Zuber, McGraw-Hill, Inc., New York, 1980. 
\bibitem{lsz}  H. Lehmann, K. Symanzik and W. Zimmermann, 
Nuovo Cimento {\bf 1}, 205 (1955). 
\bibitem{wightman} A. S. Wightman, Phys. Rev. {\bf  101}, 860 (1956).
\bibitem{lehmann} H. Lehmann, Nuovo Cimento, {\bf 10}, 579 (1958).
\bibitem{martin} A. Martin, Nuovo Cimento, {\bf 42}, 901 (1966). 
\bibitem{eden1} R. J. Eden,  High Energy Scattering, Cambridge University 
Press, 1967.
\bibitem{eden2} R. J. Eden, Rev. Mod. Phys. {\bf 43}, 15 (1971). 
\bibitem{sashanka} S. M. Roy, Phys. Rep. {\bf 5}, 125 (1972).
\bibitem{akm} G. Auberson, T. Kinishita and A. Martin, Phys. Rev. D {\bf 3}, 3185 (1971).
\bibitem{pom} I. Ya. Pomeranchuk, Zh. Eksperim. i Theor. Fiz. {\bf 34}, 725 (1958); Soviet Phys. JETP {\bf 7}, 499 (1958).
\bibitem{martinpom} A. Martin, Nuovo Cimento {\bf 39}, 704 (1965).
\bibitem{kinoshita} T. Kinoshita, in {\it Perspectives in Modern Physics}, 
Ed. R. E. Marshak, Wiley, New York, 1966, p211.
\bibitem{martinrev}  
A. Martin, Z. Phys C - Particles and Field  {\bf 15}, 185 (1982).
\bibitem{menon}  D. A. Fagundes, M. J. Menon and P. V. R. G. Silva, Braz. 
J. Phys. {\bf 42}, 452 (2012).
\bibitem{halzen}  M. Block and F. Halzen, Braz. J. Phys. {\bf 42}, 465 (2012).
\bibitem{yogi} G. Pancheri and Y.N. Srivastava , 
Eur. Phys. J. C {\bf 77}, 150 (2017).
\bibitem{leader} E. Leader, Rev. Mod. Phys. {\bf 38}, 476, 1966. 
\bibitem{roger} W. Rarita, R. J. Riddell, Jr, C. B. Chiu and R. J. N. Phillips,
Phys. Rev. {\bf 165}, 1615, 1968.
\bibitem{rp} W. Rarita and R. J. N. Phillips, Phys. Rev. Lett. {\bf 14}, 502,
1965.
\bibitem{isr_sigma_t_1} L. Baksay et al., Nucl. Phys. {\bf B141}, 1 (1978), Nucl. Phys. {\bf B148}, 538 (1979) [erratum]. 
\bibitem{isr_sigma_t_2} G. Carboni et al., Phys. Lett. B {\bf 108}, 145 (1982). 
\bibitem{isr_sigma_t_3} G. Carboni et al., Phys. Lett. B {\bf 113}, 87 (1982).
\bibitem{isr_sigma_t_4} M. Ambrosio et al. (CERN-Napoli-Pisa-Stony Brook Collaboration), Phys. Lett. B {\bf 113}, 347 (1982). 
\bibitem{isr_sigma_t_5} M. Ambrosio et al. (CERN-Napoli-Pisa-Stony Brook Collaboration), Phys. Lett. B {\bf 115}, 495 (1982). 
\bibitem{isr_sigma_t_6} G. Carboni et al., Nucl. Phys. {\bf B254}, 697 (1985). 
\bibitem{isr_sigma_t_7} N. Amos et al., Phys. Lett. B {\bf 128},343 (1983).
\bibitem{isr_sigma_t_8} D. Favart et al., Phys. Rev. Lett. {\bf 47}, 1191 (1981). 
\bibitem{isr_sigma_t_9} N. Amos et al., Nucl. Phys. {\bf B262},689 (1985). 
\bibitem{isr_sigma_t_10} N. Amos et al., Phys. Lett. B {\bf 120} 460 (1983). 
\bibitem{ua1_sigma_t} G. Arnison et al. (UA1 Collaboration), Phys. Lett. B {\bf 128} 336 (1983). 
\bibitem{ua4_sigma_t_540} T. Ekelof, in {\it Proceedings of the International Europhysics Conference on High-Energy Physics, Brighton, 1983}, p. 283; CERN-EP-83-146. 
\bibitem{ua4_sigma_t_540_2} R. Battiston et al. (UA4 Collaboration), Phys. Lett. B {\bf 117}, 126 (1982). 
\bibitem{ua4_sigma_t_541} C. Augier et al. (UA4/2 Collaboration), Phys. Lett. B {\bf 344}, 451 (1995).
\bibitem{ua4_sigma_t_546} M. Bozzo et al. (UA4 Collaboration), Phys. Lett. B {\bf 147}, 392 (1984).
\bibitem{ua5_sigma_t} J. G. Rushbrooke (for UA5 Collaboration), in {\it Proceedings of the 16th International Symposium on Multiparticle Dynamics, Israel, 1985}, p. 289; CERN-EP-85-124.
\bibitem{ua5_sigma_t_2} G. J. Alner et al. (UA5 Collaboration), Z. Phys. C {\bf 32}, 153161 (1986).
\bibitem{cdf_sigma_t_546} F. Abe et al. (CDF Collaboration), Phys. Rev. D {\bf 50}, 5550 (1994).
\bibitem{cdf_sigma_t_1800} S. N. White (for CDF Collaboration), Nucl. Phys. B Proc. Suppl. {\bf 25B}, 19 (1992).
\bibitem{e811_sigma_t_1800} C. Avila et al. (E-811 Collaboration), Phys. Lett. B {\bf 537}, 41 (2002).
\bibitem{e811_sigma_t_1800_2}C. Avila et al. (E-811 Collaboration), Phys. Lett. B {\bf 445}, 419 (1999).
\bibitem{e710_sigma_t_1020} N. A. Amos et al. (E710 Collaboration), Nuovo Cimento A {\bf 106}, 123 (1993).
\bibitem{e710_sigma_t_1800} N. A. Amos et al. (E710 Collaboration), Phys. Rev. Lett. {\bf 68}, 2433 (1992).
\bibitem{e710_sigma_t_1800_2} N. A. Amos et al. (E710 Collaboration), Phys. Rev. Lett. {\bf 63}, 2784 (1989).
\bibitem{e710_sigma_t_1800_3} N. A. Amos et al. (E710 Collaboration), Phys. Lett. B {\bf 243}, 158 (1990).
\bibitem{atlas_sigma_7k} G. Aad et al. (ATLAS Collaboration), Nucl. Phys. {\bf B889}, 486 (2014); arXiv:1408.5778 [hep-ex].
\bibitem{atlas_sigma_t_8k} M. Aaboud et al. (ATLAS Collaboration), Phys. Lett. B {\bf 761}, 158 (2016); arXiv:1607.06605 [hep-ex].
\bibitem{totem_sigma_7k} G. Antchev et al. (TOTEM Collaboration), EPL {\bf 96}, 21002 (2011).   
\bibitem{totem_sigma_7k_2} G. Antchev et al. (TOTEM Collaboration), EPL {\bf 101}, 21002 (2013).
\bibitem{totem_sigma_7k_3} G. Antchev et al. (TOTEM Collaboration), EPL {\bf 101}, 21004 (2013).
\bibitem{totem_sigma_8k} G. Antchev et al. (TOTEM Collaboration), Phys. Rev. Lett. {\bf 111}, 012001 (2013).
\bibitem{totem_sigma_t_8k_2} G. Antchev et al. (TOTEM Collaboration), Eur. Phys. J. C {\bf 76}, 661 (2016); arXiv:1610.00603 [nucl-ex].
\bibitem{totem_sigma_13k} G. Antchev et al. (TOTEM Collaboration), Eur. Phys. J. C {\bf 79}, 103 (2019); arXiv:1712.06153 [hep-ex].
\bibitem{Zbigniew} Z. Plebaniak and T. Wibig, EPJ Web of Conferences {\bf 145}, 13004 (2017). 
\bibitem{flyeye_sigma_t} R. M. Baltrusaitis, G. L. Cassiday, J. W. Elbert, P. R. Gerhardy, S. Ko, E. C. Loh, Y. Mizumoto, P. Sokolsky, and D. Steck, Phys. Rev. Lett. {\bf 52}, 1380 (1984).
\bibitem{auger_sigma_t} P. Abreu et al. (Pierre Auger Collaboration), Phys. Rev. Lett. {\bf 109}, 062002 (2012); arXiv:1208.1520 [hep-ex].
\bibitem{TelescopeArray} R. U. Abbasi et al. (Telescope Array Collaboration), Phys. Rev. D {\bf 92}, 032007 (2015); arXiv:1505.01860 [astro-ph.HE]. 
\bibitem{spmjm} S. P. Misra and J. Maharana, Phys. Rev. D {\bf 14}, 133 (1976).
\bibitem{jtalman} J. D. Talman, Special Functions: A Group Theoretic
Approach, Benjamin, New York  1969, p. 206.
\bibitem{cm} H. Corneille and A. Martin, Nucl. Phys. {\bf B115}, 163 (1976).
\bibitem{mahoux} G. Mahoux, Phys. Lett. B {\bf 65}, 139 (1976).
\bibitem{ar} G. Auberson and S. M. Roy, Nucl. Phys. {\bf B117}, 322 (1976).
\bibitem{jm} J. Maharana, Commun. Math. Phys. {\bf 58}, 195 (1978).
\bibitem{jmjkm} J. K. Mohapatra and J. Maharana, Phys. Rev. D {\bf 27}, 130, (1983).
\bibitem{isr_Amos1985} N. Amos et al., Nucl. Phys. {\bf B262}, 689 (1985).
\bibitem{ua4_dsigma_1} M. Bozzo et al. (UA4 Collaboration), Phys. Lett. B {\bf 147}, 385 (1984). 
\bibitem{ua4_dsigma_2} R. Battiston et al. (UA4 Collaboration), Phys. Lett. B {\bf 127}, 472 (1983).
\bibitem{ua4_dsigma_4} D Bernard et al. (UA4 Collaboration), Phys. Lett. B {\bf 198}, 583 (1987). 
\bibitem{cdf_sigma_el} A. Abe et al. (CDF Collaboration), Phys. Rev. D {\bf 50}, 5518 (1994).
\bibitem{e710_dsigma} N. A. Amos et al. (E710 Collaboration), Phys. Lett. B {\bf 247}, 127 (1990).
\bibitem{totem_dsigma_8k} G. Antchev et al. (TOTEM Collaboration), Nucl. Phys. {\bf B899}, 527 (2015); arXiv:1503.08111 [hep-ex].
\bibitem{totem_dsigma_13k} G. Antchev et al. (TOTEM Collaboration), Eur. Phys. J. C {\bf 79}, 861 (2019); arXiv:1812.08283 [hep-ex].
\bibitem{russian} I. M. Dremin and V. A. Nechitailo, Phys. Lett. B {\bf 720}, 177 (2013).
\bibitem{mm}  S. W. MacDowell and A. Martin, Phys. Rev. {\bf 135}, B960 (1964).
\bibitem{singh}  V. Singh, Phys. Rev. Lett. {\bf 26},530 (1971).
\bibitem{block} M. Block, L. Durand, P. Ha and F. Halzen, Phys. Rev. D {\bf 93}, 114009 (2016).

\end{thebibliography}
\end{document}